\documentstyle[11pt]{article}
\def\beq{\begin{equation}}
\def\eeq{\end{equation}}
\def\eeqlb#1{\label{#1}\eeq}
\def\bea#1#2{\renewcommand{\arraystretch}{#1}\begin{array}{#2}}
\def\eea{\end{array}}
\def\beqa#1#2{\beq \renewcommand{\arraystretch}{#1} \begin{array}{#2} }
\def\eeqa{\end{array} \eeq}
\def\eeqalb#1{\end{array} \label{#1} \eeq}
\def\seqno#1#2{\eqno(\mbox{\ref{#1}#2})}
\def\and{\hskip 5mm \mbox{and} \hskip 5mm}
\def\dps{\displaystyle}
\def\ts{\textstyle}
\def\mc#1#2#3{\multicolumn{#1}{#2}{#3}}
\def\as#1{\renewcommand{\arraystretch}{#1}}
\def\Tr{{\,\mbox{Tr}}}

\def\del{\partial}
\def\ve{\varepsilon}
\def\slt{\raisebox{-0.6ex}{$\stackrel{\ts <}{\sim}$}}
\def\hqs#1{{^{}_{Q}( #1 |}}
\def\lqs#1{{| #1 )_\ell^{}}}
\def\tp#1{\tilde\phi_{ #1 }}
\def\half{\ts\frac12}
\def\pmhalf{\pm\half}
\def\kl{k_\ell}
\def\jl{j_\ell}

\def\sQ{{s^{}_Q}}
\def\rr{\rangle\!\rangle}
\def\ll{\langle\!\langle}

\def\E{{\cal E}}
\def\I{{\cal I}}
\def\K{{\cal K}}
\def\L{{\cal L}}
  
\def\M{{\cal M}}

\def\hr{{\hat r}}
\def\vr{{\vec r}}
\def\vt{{\vec\tau}}
\def\vs{{\vec\sigma}}

\def\vo{{\vec\omega}}
\def\tdr{\vt\!\cdot\!\hr}
\def\tdo{\vt\!\cdot\!\vo}

\def\sdt{\vs\!\cdot\!\vt}

\def\vA{{\vec{A}}}
\def\Ads{\vA\!\cdot\!\vs}
\def\vL{{\vec{L}}}
\def\vS{{\vec{S}}}
\def\vI{{\vec{I}}}
\def\vK{{\vec{K}}}
\def\vKl{{\vec{K}_\ell}}
\def\vJ{{\vec{J}}}
\def\vR{{\vec{R}}}
\def\vT{{\vec{\Theta}}}
\def\tp#1{\tilde{\phi}_{#1}}
\def\tH{{\tilde{H}}}
\def\tI{{\tilde{I}}}
\def\tJ{{\tilde{J}}}
\def\tR{{\tilde{R}}}
\def\bH{\bar{H}}
\def\m{\overline{m}}
\def\MeV{\mbox{ MeV}}
\def\eff{{\mbox{\scriptsize eff}}}
\def\bff{{\mbox{\scriptsize bf}}}
\def\rot{{\mbox{\scriptsize rot}}}
\def\adj{{\mbox{\scriptsize adj}}}
\def\gF{gF^\prime(0)}
\def\vsl{{v \!\!\!/}}
\def\ie{{\it i.e.\,}}
\def\eg{{\it e.g.\,}}

\def\etal{{\it et. al.\,}}
\def\Pcs{P_{\bar{c}s}^{}}

\def\Pc{P_{\bar{c}}^{}}
\def\Pb{P_{\bar{b}}^{}}
\def\fbs#1#2#3{{Few-Body Sys. }{\bf #1}{ (#2) }{#3}}

\def\npa#1#2#3{{Nucl. Phys. }{\bf A#1}{ (#2) }{#3}}
\def\npb#1#2#3{{Nucl. Phys. }{\bf B#1}{ (#2) }{#3}}

\def\plb#1#2#3{{Phys. Lett. }{\bf B#1}{ (#2) }{#3}}
\def\prl#1#2#3{{Phys. Rev. Lett. }{\bf #1}{ (#2) }{#3}}

\def\prd#1#2#3{{Phys. Rev. }{\bf D#1}{ (#2) }{#3}}

\def\zpa#1#2#3{{Z. Phys. }{\bf A#1}{ (#2) }{#3}}

\begin{document}
\begin{titlepage}

\rightline{SNUTP-94/06 (Revised)}

\rightline{April, 1994}

\vskip 0.5cm

\begin{center}
{\LARGE \bf Pentaquark Exotic Baryons in the Skyrme Model}
\vskip 1.5cm
{\large Yongseok OH} \vskip 3mm
{\it Department of Physics, National Taiwan University}\\
{\it Taipei, Taiwan 10764, R.O.C.} \vskip 5mm
{\it and} \vskip 5mm
{\large Byung-Yoon PARK} \vskip 3mm
{\it Department of Physics, Chungnam National University}\\
{\it Deajeon 305--764, Korea} \vskip 5mm
{\it and} \vskip 5mm
{\large Dong-Pil MIN}\vskip 3mm
{\it Center for Theoretical Physics and Department of Physics}\\
{\it Seoul National University, Seoul 151--742, Korea}\\
\vskip 2.5cm
{\large\bf ABSTRACT}
\begin{quotation}
We investigate the pentaquark($P$) exotic baryons as
soliton-antiflavored heavy mesons bound states in the limit of
infinitely heavy meson mass. Our approach respects the chiral
symmetry as well as the heavy quark symmetry. The results reveal
a possibility for the loosely bound non-strange $P$-baryon(s).
\end{quotation}
\end{center}
\end{titlepage}
%------------------------ Introduction --------------------------
A pentaquark ($P$) baryon \cite{Lipkin,GSR} is an exotic baryon
predicted in quark models, which consists of a heavy antiquark
$\bar{Q}$($\bar{c}$ or $\bar{b}$) and four light quarks such as
ordinary $q^{}_0$($u,d$) quarks and $s$-quark. Its stability is
provided by a gain in the hyperfine interaction energy stemming
from one gluon exchange just as in the $H$-dibaryon \cite{Jaffe}.
Lipkin \cite{Lipkin} and Gignoux \etal \cite{GSR} show that, in the
limit of the infinite $c$-quark mass and the exact $SU(3)_F$
symmetry for the light quarks, a strange anti-charmed baryon
$\Pcs$ $(\bar{c}sq_0q_0q_0)$ is stable against the decays into
$\Lambda D$ or $N D_s$. The binding energy of $\Pcs$ over the
$q_0q_0s$-$\bar{Q}q_0$ or $q_0q_0q_0$-$\bar{Q}s$ system is
$\sim 150$~MeV, which becomes down to $\sim 85$ MeV if included a
realistic $SU(3)_F$ symmetry breaking \cite{KZ}. Furthermore, the
system becomes unbound if quark motions are taken into
account \cite{FGRS,ZR}.

In the Skyrme model, the $P$-baryon can be described by a
soliton-antiflavored heavy meson bound state (if any).
Recently, Riska and Scoccola (RS) \cite{RS93} reported a few
{\em non-strange} $P$-baryon states such as
$\Pc(\bar{c}q_0 q_0 q_0 q_0)$ and $\Pb(\bar{b} q_0 q_0 q_0 q_0)$
in the extended bound state approach \cite{CK} of the Skyrme model.
The lowest $\Pc$ and $\Pb$ are the isosinglet ($i$=0) and spin
doublet ($j$=$\frac12$) states, whose binding energies with respect
to the $ND$ and $NB$ threshold are estimated to be
$110 \sim 190$~MeV and $0.7 \sim 1.0$~GeV, respectively.
It is quite a remarkable result compared with the quark model
by which a nonstrange anti-charmed $P$-baryon has no sufficient
symmetry to be stable via the hyperfine interactions.

Following the traditional collective coordinate quantization
procedure \cite{CK}, RS obtained the mass formula for such
nonstrange $P$-baryon, $m_P^{}$, as
\beq
m_P^{} = M_{sol} + \omega
+ \frac{1}{2\I}\{cj(j+1) + (1-c)i(i+1)
+c(c-1)k(k+1)\}.
\eeqlb{MFRS}
Here, $M_{sol}$ and $\I$ are the soliton mass and its moment of inertia
with respect to the collective isospin rotation, $\omega$ and $k$ are the
eigenenergy and the grand spin of the bound state for the antiflavored
heavy meson, and $i$($j$) are the isospin(spin) of the $P$-baryon.
It has been understood that the ``hyperfine constant", $c$, should vanish
in the heavy meson mass limit so that the masses do not depend on
the spin, while their numerically obtained hyperfine constants
for the bound $B$=+1 heavy mesons do not vanish. The worse is that
they can be negative, which implies that the state with higher spin has
the lower mass. On the other hand, the heavy quark spin symmetry
does not necessarily lead to such an independence of  mass on the spin.
It just implies that the hadrons containing a single heavy quark
(or a single antiquark) come in degenerate doublets of total spin
$j$=$\jl\pm\half$ with $\jl$ being the spin of the light degrees of
freedom, while the traditional mass formula (\ref{MFRS}) is not
convenient to see this kind of symmetry explicitly. Consequently, the
work of RS does not respect the heavy quark symmetry. \cite{HQS}

In addition, their way of treating the heavy vector mesons is not
consistent with the heavy quark symmetry, either. \cite{HB} They
integrate out the heavy vector meson fields in favor of the pseudoscalar
ones, which may be guaranteed only when the former are sufficiently
heavier than the latter. Since the vector mesons $D^*$ and $B^*$ are
only a few percent heavier than their pseudoscalar partners $D$ and
$B$, one should treat the heavy vector mesons on the same footing as
the heavy pseudoscalar mesons. We are to re-examine the existence of
nonstrange $P$-baryons in the Skyrme model by correcting these defects
of the work of ref. \cite{RS93}. In this paper, as a first step, we
report briefly our estimation on the binding energy in the infinitely
heavy meson mass limit with a theory respecting the heavy quark symmetry
and the chiral symmetry.

%------------------- Effective Lagrangian -------------------------
Consider a system of Goldstone pions ($\pi^+$, $\pi^0$ and $\pi^-$)
and $j^\pi$=$0^-$ and $1^-$ heavy mesons containing a sufficiently heavy
antiquark $\bar{Q}$ and a light quark $q$. The dynamics of the system
is governed by the $SU(2)_L\times SU(2)_R$ chiral symmetry and the
heavy quark symmetry \cite{HQS}. As the mass of the heavy constituent
becomes sufficiently larger than the typical scale of strong
interactions, its spin decouples to the rest so that the $0^-$ and
$1^-$ heavy mesons have the same mass and furthermore the dynamics of
the system is independent of the heavy constituent. \cite{HQS}
The incorporation of the heavy quark symmetry is facilitated by
representing the $0^-$ and $1^-$ heavy mesons by a $4\times4$
(isodoublet) matrix field $H(x)$ as
\beq
H = \frac{1-\vsl}{2} \left( \Phi_v \gamma_5
   - \Phi^*_{v\mu}\gamma^\mu \right).
\eeqlb{H}
Here, $\gamma$'s are the $4\times4$ Dirac matrices and $\vsl$ denotes
$v_\mu\gamma^\mu$. And $\Phi_v$ and $\Phi_{v\mu}^*$, respectively,
represent the heavy pseudoscalar field and heavy vector fields moving
with a four velocity $v_\mu$. As inferred from the $q\bar{Q}$ structure,
under the heavy quark spin rotation, $H$ transforms
\beq
H \rightarrow H S^{-1},
\eeq
with $S\in SU(2)_v$ (the heavy quark spin symmetry group
boosted by the velocity $v$).

As for the system of Goldstone pions, the chiral symmetry is
realized in a nonlinear way by a $2\times2$ unitary matrix
\beq
\Sigma = \exp \left(\frac{i}{f_\pi}\vt\cdot\vec{\pi} \right),
\eeq
which transforms, under an $SU(2)_L\times SU(2)_R$ transformation,
\beq
\Sigma \rightarrow L \Sigma R^\dagger,
\eeq
with global transformations $L\in SU(2)_L$ and $R\in SU(2)_R$.
Here, $f_\pi$ is the pion decay constant. In terms of $\Sigma$,
the interactions among the Goldstone bosons are described by the
lagrangian density
\beq
\L_M = \frac{f_\pi^2}{4}\Tr(\del_\mu \Sigma^\dagger \del^\mu \Sigma)
+ \cdots ,
\eeq
where terms with higher derivatives are denoted by the ellipsis.
With a suitable stabilizing term provided, the nonlinear lagrangian
$\L_M$ supports a classical soliton solution
\beq
\Sigma_0(\vr)=\exp(i\tdr F(r)),
\eeqlb{SS}
with the profile function satisfying the boundary conditions
$F(0)=\pi$ and $F(r)\stackrel{r\rightarrow\infty}{\longrightarrow} 0$.
To this purpose, among the higher derivative terms, we include the
conventional Skyrme term
\beq
\L_{\mbox{\tiny SK}}=\frac{1}{32e^2}\Tr[\Sigma^\dagger\del_\mu \Sigma,
\Sigma^\dagger\del_\nu \Sigma]^2,
\eeq
into $\L_M$ with a dimensionless parameter $e$.

Let $\xi\equiv \Sigma^{1/2}$ be a redefined matrix
which transforms under an $SU(2)_L\times SU(2)_R$ as
\beq
\xi \rightarrow L \xi U^\dagger = U \xi R^\dagger,
\eeqlb{ct1}
with a special unitary matrix $U$ depending on $L$, $R$ and the
Goldstone fields. And let's assign $H(x)$ a transformation rule as
\beq
H \rightarrow U H.
\eeqlb{ct2}
Then, to the leading order in the derivatives on the Goldstone boson fields,
a {\em ``heavy-quark-symmetric"} and {\em ``chirally-invariant"}
lagrangian can be written as \cite{HQEF}
\beq
\L_{HQS}=\L_M - i v_\mu \Tr(\bH (\del^\mu + V^\mu) H)
 + g \Tr (\bH\gamma^\mu \gamma_5 A_\mu H),
\eeqlb{LHQS}
with a universal coupling constant $g$ for the
$\Phi^*\Phi\pi$ and $\Phi^*\Phi^*\pi$ interactions.
Vector fields $V_\mu(x)$ and axial vector fields $A_\mu(x)$
are defined and transform under the chiral transformation as
\beqa{1.2}{l}
V_\mu = \frac12 (\xi^\dagger \del_\mu \xi + \xi \del_\mu \xi^\dagger)
\rightarrow U V_\mu U^\dagger + U \del_\mu U^\dagger,\\
A_\mu = \frac{i}2 (\xi^\dagger \del_\mu \xi - \xi \del_\mu \xi^\dagger)
\rightarrow U A_\mu U^\dagger.
\eeqa

In our model lagrangian (\ref{LHQS}), we have three parameters; the pion
decay constant $f_\pi$, the Skyrme parameter $e$ and the pion-heavy meson
coupling constant $g$. Empirical value of $f_\pi$ is 93 MeV. The
nonrelativistic quark model provides a naive estimation for the value
of $g$ as $g=-\frac34$ \cite{Yan} and, in case of $Q$=$c$, the
experimental upper limit \cite{ACCMOR} of the $D^*$ width ($\sim$ 131 keV)
implies  $|g|^2\slt 0.5$ when combined with the
$D^{*+}\rightarrow D^+\pi^0$ and $D^{*+}\rightarrow D^0\pi^+$ branching
ratios \cite{CLEO}. We will take them however as free parameters and
adjust them to produce experimentally observed heavy baryon masses.

%----------------------- Soliton-Heavy Meson Bound State -----------------

Our main interest is the heavy meson bound states to the static
potentials provided by the baryon-number-one soliton configuration
(\ref{SS}); which are explicitly
\beqa{1.2}{l}
V^\mu=(V^0,\vec{V})=(0,i\upsilon(r)\hr\times\vt), \\
A^\mu=(A^0,\vec{A})=(0,\frac12(a_1(r)\vt + a_2(r)\hr\tdr)),
\eeqa
with
\beq
\upsilon(r)=\frac{\sin^2(F/2)}{r}, \hskip 5mm
a_1(r)=\frac{\sin\!F}{r} \and
a_2(r)=F^\prime - \frac{\sin\!F}{r}.
\eeq

In the rest frame where $v^\mu=(1,\vec{0})$, the equation of motion
for the eigenmodes $H_n(\vr)$ of the $H$-field with the eigenvalue
$\ve_n$ can be read off as
\beq
\ve_n H_n(\vr) = - g \vs\cdot\vec{A} H_n(\vr)  ,
\eeqlb{Heq}
where we have used the relation
$\vec{\gamma}\gamma_5 H(x)= \vec{\sigma} H(x)$,
and $n$ denotes a set of quantum numbers to classify the eigenmodes.
The ``hedgehog" configuration (\ref{SS}) correlates the isospin and the
angular momentum, while the heavy-quark symmetry implies the heavy quark
spin decoupling. Thus, the equation of motion is invariant under the parity
operation, under the heavy quark spin rotation and under the simultaneous
rotations in the isospin space, {\em light-quark} spin space and
ordinary spaces. Let $\vL$, $\vS_\ell$, $\vS_Q$ and $\vI_h$ be the orbital
angular momentum, light quark spin, heavy quark spin and isospin
operators of the heavy mesons, respectively. And $Y_{\ell m}(\hr)$,
$\lqs{\pmhalf}$, $\hqs{\pmhalf}$, $\tp{\pmhalf}$ be their corresponding
eigenfunctions or eigenstates, respectively. The simultaneous rotations
mentioned above are generated by the ``{\em light-quark grand spin}"
operator defined as\footnote{By the subscript $\ell$, we distinguish
$\vec K_\ell$ from the traditional grand spin operator used in the bound
state approach in the Skyrme model; \ie,
$\vec{K}$($=\vL+\vS+\vI_h$)  with $\vS$($=\vS_\ell+\vS_Q$)
being the spin operator of the heavy mesons.}
\beq
\vKl=\vL+\vS_\ell+\vI_h.
\eeq
Then, the eigenmodes of the heavy meson can be classified by the
third component of the heavy quark spin $\sQ$, the grand spin and its
third component $(\kl,k_3)$ and the parity $\pi$. The set of quantum
numbers will be denoted by $n=\{\kl,k_3,\pi,\sQ\}$.

The situation is very similar to obtaining the eigenmodes of the confined
quarks in the chiral bag model \cite{BYP}. We start with the construction
of the eigenfunctions of the grand spin and the heavy quark spin by
combining the direct products of the four angular momentum eigenstates,
$Y_{\ell m_\ell}$, $\tp{\pmhalf}$, $\lqs{\pmhalf}$ and
$\hqs{\pmhalf}$: \cite{OPM1}
\beq
\K^{(i)}_{\kl k_3 \sQ} = \dps \sum_{m_s,m_t}
(\ell^{(i)},m_\ell,\half,m_t|\lambda^{(i)},m_\ell+m_t)
(\lambda^{(i)},m_\ell+m_t,\half,m_s|\kl,k_3) \;
Y_{\ell m_\ell}(\hr) \tp{m_t} {\lqs{m_s} \, \hqs{\sQ}}, \\
\eeq
with the help of Clebsch-Gordan coefficients
$(\ell_1,m_1,\ell_2,m_2|\ell,m)$. Here, we first combine the orbital
angular momentum and the isospin ($\vec{\lambda}=\vL+\vI_h$) and then
combine the light quark spin. For a given $\kl(\neq 0)$, we have four
$\K^{(i)}_{\kl k_3 \sQ}$ depending on $\ell$ and $\lambda^{(i)}$.
(See Table~1.)
\begin{table} %----------- Table 1 --------------
\begin{center}
{\bf Table 1} : Four $\K^{(i)}_{\kl k_3 s_Q}$-basis.
\vskip 3mm
\begin{tabular}{ccc}
\hline
$i$ & $\lambda$ & $\kl$ \\
\hline
1 & $\ell+\half$ & $\lambda-\half=\ell$ \\
2 & $\ell-\half$ & $\lambda+\half=\ell$ \\
3 & $\ell-\half$ & $\lambda-\half=\ell-1$ \\
4 & $\ell+\half$ & $\lambda+\half=\ell+1$ \\
\hline
\end{tabular}\end{center}
\end{table} %------------- End of Table 1 --------------

In terms of these $\K^{(i)}_{\kl k_3 \sQ}$, the heavy meson wavefunction
can be written as
\beqa{1.5}{ll}
\dps H_n(\vr) = \sum_{i=1,2} h^{(i)}_{\kl}(r)
\K^{(i)}_{\kl k_3 \sQ} , &
\mbox{for $\pi=-(-1)^{\kl}$ states}, \\
\dps H_n(\vr) = \sum_{i=3,4} h^{(i)}_{\kl}(r)
\K^{(i)}_{\kl k_3 \sQ} , &
\mbox{for $\pi=+(-1)^{\kl}$ states},
\eeqa
with the radial functions $h^{(i)}_{\kl}(r)$. Note that $i$=1,2 states
and $i$=3,4 states are decoupled due to parity.

We assume that the heavy meson and the soliton are infinitely heavy,
in which case the heavy meson is just sitting at the center of the soliton
where the potential has the lowest value. That is, in the heavy mass
limit, all the radial functions $h^{(i)}_{\kl}(r)$ can be approximated as
\beq
h^i_{\kl}(r) = \alpha_i f(r),
\eeqlb{hr}
with a constant $\alpha_i$ and a function $f(r)$ which is strongly
peaked at the origin and normalized as
$ \int^\infty_0 r^2dr |f(r)|^2 = 1 $. Then, the problem is reduced
to solving the secular equation
\beq
\sum_{j} \M_{ij} \alpha_j = - \ve \alpha_i,
\hskip5mm\mbox{ ($i,j$=1,2 or 3,4)}
\eeqlb{SecEq}
where the matrix elements $\M_{ij}(i,j$=1,2 or 3,4) are defined as
\beq
\M_{ij} = -\ts\frac12\gF \dps
\int\!d\Omega \Tr\left\{ \bar{\K}^{(i)}_{\kl k_3 \sQ}
(\tdr)[(\sdt)](\tdr)
\K^{(j)}_{\kl k_3 \sQ} \right\},
\eeqlb{M}
with $ \bar{\K} = \gamma^0 \K^\dagger \gamma^0$. The minus sign of the
energy in eq. (\ref{SecEq}) comes from the normalization of the basis
states $\K$. We have used that $F(r) = \pi + F^\prime(0) r + O(r^3)$ near
the origin so that $a_1(r)\sim -F^\prime(0) + O(r^2)$ and
$a_2(r)\sim 2F^\prime(0) + O(r^2)$ and the identity
$(2\vs\cdot\hr\vt\cdot\hr-\vs\cdot\vt)
=(\vt\cdot\hr)(\vs\cdot\vt)(\vt\cdot\hr)$.

For a given set of quantum numbers for $\kl$, $k_3$ and $\sQ$, we have four
eigenstates\footnote{In case of %----------- Footnote ----------
$\kl=0$, we have two eigenstates;
$$\renewcommand{\arraystretch}{1.3}\begin{array}{ll}
\ve=-\frac12\gF; & \K^{(1)}_{0 0 s_Q}(\hr),  \\
\ve=+\frac32\gF; & \K^{(3)}_{0 0 s_Q}(\hr).
\end{array}$$ }: %------------------------------- Footnote end ------
\beqa{1.7}{ll}
%---------------------------------------------
\ve=-\ts\frac12\gF; &
\K^{(1)}_{\kl k_3 s_Q}, \K^{(2)}_{\kl k_3 s_Q},
\K^{(+)}_{\kl k_3 s_Q} (=
\sqrt{\frac{\kl}{2\kl+1}} \K^{(3)}
 + \sqrt{\frac{\kl+1}{2\kl+1}} \K^{(4)}),  \\
%---------------------------------------------
\ve=+\ts\frac32\gF; &
\K^{(-)}_{\kl k_3 s_Q} (=
\sqrt{\frac{\kl+1}{2\kl+1}} \K^{(3)}
 - \sqrt{\frac{\kl}{2\kl+1}} \K^{(4)}).
\eeqa
Since $gF^\prime(0)>0$ (in case of baryon-number-one soliton solution),
we have {\em three} bound states of the heavy mesons carrying
antiflavor ($C$=$-$1 or $B$=+1) with a binding energy $\frac12\gF$.
The unbound state with eigenenergy $+\frac32\gF$ corresponds to the
bound state of the heavy meson with $C=+1$ or $B=-1$, whose eigenstates
are given by negative energy eigenstates in our approach. It is
interesting to note that a $\kl$ leads to two grand spins; \ie,
$k=\kl\pm\half$ (unless $\kl=0$). Thus, if one works with the grand spin
$\vK$ (instead of $\vKl$) along with the traditional bound state approach,
the heavy quark spin symmetry implies that the eigenstates come in by
degenerate doublets with grand spin $k=\kl\pm\half$. When the heavy mesons
have finite masses, these degeneracies should be removed. In other words,
as the heavy meson masses increase, their eigenstates approach each other
and become degenerate in the infinite heavy meson mass limit.
It plays a nontrivial role in the quantization procedure.

In case of a typical soliton solution stabilized by the Skyrme term,
when parameters are fixed as $f_\pi$=64.5 MeV and $e$=5.45
for the soliton to fit the nucleon and Delta masses \cite{ANW},
$F^\prime(0)$ amounts to $\sim -0.70$ GeV. With $g=-0.75$ that the
nonrelativistic quark model predicts, the binding energy of the normal
heavy mesons and anti-flavored heavy mesons to the soliton are estimated
as $\frac32\gF\sim 0.79$ GeV and $\frac12\gF\sim 0.26$ GeV,
respectively. Comparing it with that of ref. \cite{RS93}, one can see that
the binding energy for the bound antiflavored heavy meson is reduced
by a factor of one half and more. It should be mentioned further that in
ref. \cite{RS93} the binding energy increases as the heavy meson masses
and also that our results are obtained with infinite heavy meson masses.

However, the degeneracy in $\kl$ is an artifact originated from the
approximation (\ref{hr}) on the radial function $h^{(i)}_{\kl}(r)$.
In general, when the heavy meson's kinetic term is taken into account,
the radial function feels the centrifugal potential
$\ell_{\eff}(\ell_\eff+1)/r^2$ near the origin so that
it behaves as $h^{(i)}_{\kl}\sim r^{\ell_\eff}$.
Here, $\ell_\eff$ is the ``effective" angular momentum \cite{CK},
which is related with $\ell$ as
\beq
\ell_\eff = \left\{
\renewcommand{\arraystretch}{1}\begin{array}{ll}
 \ell+1, & \mbox{if $\lambda_i=\ell+\frac12$}, \\
 \ell-1, & \mbox{if $\lambda_i=\ell-\frac12$}.
\end{array} \right.
\eeq
Due to the vector potential $\vec{V}$ ($\sim i(\hr\times\vec\tau)/r$,
near the origin) the singular structure of the {\it covariant}
derivative squared $\vec{D}^2=(\vec{\nabla}+\vec{V})^2$ is altered to
$\ell_\eff(\ell_\eff+1)/r^2$ from the usual form of $\ell(\ell+1)/r^2$,
which results from $\vec{\nabla}^2$. Thus, only those states with
$\ell_\eff=0$ can have strongly peaked radial function and the
degeneracies will split such that the states with higher $\ell_\eff$
have higher energy. Note that $\ell_\eff=0$ can be
achieved only when $\ell=1$.

It should be mentioned here that the wavefunctions appear in an ill-defined
form at the origin\footnote{We thank to the
referee for pointing out this matter.}.
For example, the grand spin eigenfunction $K^{(2)}_{\kl=1, k_3,s_Q}$
is explicitly
\begin{equation}
\begin{array}{l}
K^{(2)}_{1,+1,s_Q} = (\vec{\tau}\cdot\hat{r}) \tilde{\phi}_{+\frac12}^{}
\, \lqs{+\frac12} \, \hqs{\sQ}, \\
K^{(2)}_{1,0,s_Q} = \frac{1}{\sqrt2} (\vec{\tau}\cdot\hat{r})
\left\{ \tilde{\phi}_{+\frac12}^{}\, \lqs{-\frac12} \, \hqs{\sQ}
+ \tilde{\phi}_{-\frac12}^{}\, \lqs{+\frac12} \, \hqs{\sQ}
 \right\}, \\
K^{(2)}_{1,-1,s_Q} = (\vec{\tau}\cdot\hat{r}) \tilde{\phi}_{+\frac12}^{}
\, \lqs{+\frac12} \, \hqs{\sQ}.
\end{array}
\end{equation}
Since the radial function $h(r)$ does not vanish at the origin;
{\it i.e.,} $r \sim r^{\ell_\eff}$ with
$\ell_\eff=0$, $(\vec{\tau}\cdot\hat{r})$
in $K^{(2)}_{1k_3s_Q}$ leads an ill-defined wavefunction.
Together with $\xi=\exp(i\vec{\tau}\cdot\hat{r} F(r)/2)$ which is
also ill-defined at the origin, such ill-defined wavefunctions
do not cause any problem in evaluating the physical quantities.
It is entirely due to our convention for the representation
of chiral symmetry. We have adopted, so-called the $\xi$-basis
which has a simple transformation rule for the parity, while
one would have well-defined wavefunctions in the $\Sigma$ basis.
(See ref. \cite{xiSigma} for further details.)

To endow correct quantum numbers such as spin and isospin to the
soliton-heavy meson bound system, we quantize the zero modes
associated with the invariance under simultaneous $SU(2)$ rotation
of the soliton configuration together with the heavy meson fields.
We introduce the $SU(2)$ collective variables $C(t)$ as
\beq
\xi(\vr,t) = C(t) \xi^{}_0(\vr) C^\dagger(t), \and
H(\vr,t) = C(t) H_{\bff}(\vr,t).
\eeqlb{CV}
Here, $H_{\bff}$ refers to the heavy meson field in the body-fixed frame,
while $H(\vr,t)$ refers to that in the laboratory frame. Substitution of
eq. (\ref{CV}) into eq. (\ref{LHQS}) leads us to the lagrangian (in the
reference frame where the heavy meson is at rest in space but rotating in
isospin space)
\beqa{1.5}{l}
\ts L^{\rot}
= - M_{sol}
  + \dps\int\!d^3r \left\{ -i\Tr(\bH_\bff\del_0 H_\bff)
  + g\Tr(\bH_\bff \Ads H_\bff )  \right\} \\
\hskip 1cm
+ \half\I\omega^2
  + \dps\int\!d^3r \ts \frac12
\Tr \left\{ \bH_\bff \ts\frac12(\xi^\dagger \vec \tau \cdot \vec
    \omega \xi + \xi \vec\tau \cdot \vec\omega \xi^\dagger) H_\bff
   \right\},
\eeqalb{Lcol}
where we have kept terms up to $O(m^0_QN_c^{-1})$.
The ``angular velocity", $\vo$,  of the collective
rotation is defined by $ C^\dagger\del_0 C \equiv \half i\tdo. $

The lagrangian (\ref{Lcol}) leads us to the Hamiltonian as
\beq
\tH = M_{sol} - g\!\int\!d^3r\,
\Tr(\bH_\bff\,\Ads\,{H}_{\bff})
+ \frac{1}{2\I}(\vR-\vT(\infty))^2 ,
\eeqlb{Hcol}
where the rotor spin $\vR$ is canonical conjugate
to the collective variables $C(t)$:
$$ R_a \equiv \frac{\delta L^{\rot}}{\delta\omega^a}
= \I \omega_a + \Theta_a(\infty),
\seqno{Hcol}{a} $$
with $\vT(\infty)$ defined as
$$ \vT(\infty) \equiv
+ \ts \frac12 \dps \int\! d^3r\Tr \left\{ \bH_\bff \, \ts
\frac12(\xi^\dagger \vt \xi
+ \xi \vt \xi^\dagger) \, H_\bff\right\}.
\seqno{Hcol}{b} $$

With the collective variable introduced as in eq. (\ref{CV}), the
isospin of the fields $U(x)$ and $H(x)$ is entirely attributed to $C(t)$
and the isospin operator can be written in terms of the rotor spin as
\beq
I_a = \half\Tr(\tau_a C \tau_b C^\dagger)
( \I \omega_b + \Theta_b(\infty) ) = D^\adj_{ab}(C) R_b,
\eeq
with $D^\adj_{ab}(C)$ being the $SU(2)$ adjoint representation associated
with the collective variables $C(t)$. Furthermore, with the help of the
$K$-symmetry in the solution, one can easily show that the spin of the
$H_{\bff}$ is the grand spin; that is, the isospin of the $H$-field is
transmuted into the part of the spin in the body-fixed frame. The spin
of the soliton-heavy meson bound system is obtained as
\beq
\vJ=\vR + \vec{K}^{\bff},
\eeq
with the grand spin $\vec{K}^{\bff}$ of the heavy meson fields in the
body-fixed frame. Finally, the
heavy-quark spin symmetry of the lagrangian under the transformation
$H(x) \rightarrow H(x) S^{-1} = C(t) (H_{\bff}(x) S^{-1})$ has nothing
to do with the collective rotations. Because of this heavy-quark spin
decoupling, it is convenient to proceed with the spin operator $\vJ_\ell$
for the light degrees of freedom in the soliton-heavy meson bound system
defined as
\beq
\vJ_\ell = \vJ - \vS_Q = \vR + \vec{K}_\ell^{\bff}.
\eeqlb{Jell}

Upon canonical quantization, the collective variables become the quantum
mechanical operators; the isospin ($\vI$), the spin ($\vJ_\ell$) and the
rotor spin ($\vR$) discussed so far become the corresponding operators
$\tI_a$, $\tJ_{\ell,a}$ and $\tR_a$, respectively. We distinguish those
operators associated with the collective coordinate quantization by using
a tilde on them. Let's denote the eigenstates of the rotor-spin operator
$\tR_a$ as $|i;m_1,m_2\}$ $(m_1,m_2=-i,-i+1, \cdots, i)$:
\beqa{1.2}{c}
\tR^2 |i;m_1,m_2\} = i(i+1)|i;m_1,m_2\}, \\
\tR_3 |i;m_1,m_2\} = m_2 |i;m_1,m_2\}, \\
\tI_3 |i;m_1,m_2\} = m_1 |i;m_1,m_2\}.
\eeqa
Then, the eigenstates $|i,i_3;j_\ell,j_{\ell,3};\sQ\rr$ of the operators
$\tI_a$ and $\tJ_{\ell,a}$ with their corresponding quantum numbers
$i,i_3$ (isospin) and $j_\ell,j_{\ell,3}$ (spin of the light degrees
of freedom) are given by the linear combinations of the direct product
of the rotor spin eigenstate $|i;m_1,m_2\}$ and the single particle Fock
state $|n\rangle$:
\beq
|i,i_3;j_\ell,j_{\ell,3};\sQ\rr_{a}^{}
=\sum_{m} (i,j_{\ell,3}-m,k_\ell^a,m|j_\ell,j^{}_{\ell,3})
|i;i_3,j_{\ell,3}-m\} |\kl,m,\sQ\rangle_a^{}.
\eeqlb{BS}
One may combine further the heavy quark spin and the spin of the light
degrees of freedom to construct the states with a good total spin, which
is not necessary however. Note that, in the infinite heavy quark mass
limit, $(j_\ell,j_{\ell,3})$ themselves are good quantum numbers
of the heavy hadrons together with the heavy quark spin due to the heavy
quark symmetry. For a given set of $(i,j^\pi_\ell)$, there can be more
than one state depending on which Fock state $|n\rangle$ is involved in
the combination (\ref{BS}). We will distinguish them by using a
sequential number, $a$(=1,2$\cdots$); \eg, $|i,j_\ell^\pi\rr_1$.
Here again, to shorten the expressions, we will not specify
such quantum numbers as $i_3$, $j_3$ and $\sQ$ unless necessary.
In Table~2, we list a few $|i,j^\pi_\ell\rr$ states resulting from the
soliton-antiflavored heavy meson bound system.
Here, we have included only the integer rotor spin states so that the
combined states can have a half-integer spin ($j=j_\ell\pmhalf$).
%------------------------- Table 2 --------------------------
\begin{table}
\begin{center}\as{1.5}
{\bf Table 2 } : $|i,j_\ell^\pi\rr$ states for the $P$-baryons.
\vskip 3mm
\begin{tabular}{ccllcc}
\hline
 $i$ & $j_\ell^\pi$ & $\dps\bea{1}{l} |n\rangle \eea$ &
$\dps\bea{1}{l} |i,j^\pi_\ell\rr_i \eea$ & $\ve$
& $j$ \\ \hline
\hline %------------------------P i=0 j_\ell=0^- ----------------
 0 & $0^-$ & $\dps\bea{1}{l} |0\rangle_1 \eea$ &
$\dps\bea{1}{l} |0,0^-\rr \eea$ & $-\frac12\gF$ & $\half$   \\
\hline %------------------------P i=0 j_\ell=1^- ----------------
 0 & $1^-$ & $\dps\bea{1}{l} |1\rangle_+ \eea$ &
$\dps\bea{1}{l} |0,1^-\rr \eea$ & $-\frac12\gF$ & $\half,\frac32$  \\
\hline %------------------------P i=1 j_\ell=1^- ----------------
 1 & $1^-$ & $\dps\bea{1}{l} |0\rangle_1 \\ |1\rangle_+ \\
 |2\rangle_1 \\ |2\rangle_2 \eea$ &
$\dps\bea{1}{l} |1,1^-\rr_1 \\ |1,1^-\rr_2 \\
|1,1^-\rr_3 \\ |1,1^-\rr_4 \eea$ & $-\frac12\gF$ &
$\half,\frac32$  \\ \hline
\hline %------------------------P i=0 j_ell=1^+ -----------------
 0 & $1^+$ &
$\dps\bea{1}{l} |1\rangle_1 \\ |1\rangle_2 \eea$ &
$\dps\bea{1}{l} |0,1^+\rr_1 \\ |0,1^+\rr_2 \eea$
& $-\frac12\gF$ & $\half,\frac32$  \\
\hline %------------------------P i=1 j_ell=0^+ -----------------
 1 & $0^+$ &
$\dps\bea{1}{l} |1\rangle_1 \\ |1\rangle_2 \eea$ &
$\dps\bea{1}{l} |1,0^+\rr_1 \\ |1,0^+\rr_2 \eea$
& $-\frac12\gF$ & $\half$  \\
\hline %------------------------P i=1 j_ell=1^+ -----------------
 1 & $1^+$ &
$\dps\bea{1}{l} |1\rangle_1 \\ |1\rangle_2 \\ |2\rangle_+ \eea$ &
$\dps\bea{1}{l} |1,1^+\rr_1 \\ |1,1^+\rr_2 \\ |1,1^+\rr_3 \eea$
& $-\frac12\gF$ & $\half,\,\frac32$  \\ \hline
\end{tabular}
\end{center}
\end{table}
%-------------------------- Table 2 End ------------------------

The physical $P$-baryons under consideration appear as the eigenstates
of the Hamiltonian $\tH$. In case that we have only a single state
$|i,j_\ell\rr$ for a given quantum numbers $(i,j_\ell^\pi)$, it is the
mass eigenstate and then the mass (modulo the heavy meson masses) of
the corresponding baryon is simply obtained by evaluating the
expectation value of the Hamiltonian with respect to the state:
\beq
M_{(i,j_\ell)} = M_{sol} + \ve_n + \frac{1}{2\I}
\{(1-c)i(i+1)+cj_\ell(j_\ell+1)-c\kl(\kl+1) + \ts\frac34\},
\eeqlb{E1}
where $\ve_n$ and $\kl$ are the eigenenergy and the light-quark grand
spin of the heavy meson bound state involved in the combination of
$|i,j_\ell^\pi\rr$. We have used that the expectation values of the
operators $\vT(\infty)$ and $\vT^2(\infty)$ with respect to the same
single particle Fock state $|n\rangle$ are
$$
\langle n|\vT(\infty)|n\rangle
= - c \langle n|\vK_\ell|n\rangle,
\seqno{E1}{a}
$$
with a constant $c$, and for all $|n\rangle$
$$
\langle n|\vT^2(\infty)|n\rangle = \ts\frac34.
\seqno{E1}{b}
$$
Note that the mass formula (\ref{E1}) respects the heavy quark symmetry,
{\em regardless of the $c$-value}, and that eq. (\ref{E1}b) is different
from what would have been obtained by using the approximation of the
traditional bound state approaches \cite{CK}
\beq
 \langle n|\vec\Theta^2(\infty)|n\rangle
  \approx |\langle n|\vT(\infty)|n\rangle|^2 = c^2 \kl(\kl+1).
\eeq

If we have more than one state, say $|i,j_\ell^\pi\rr_{a}(a$=$1,2,\cdots$),
the situation is little bit more complicated. In general, each state alone
cannot be an eigenstate of the Hamiltonian. Since they are degenerate
up to the order $N_c^0$, the mass cannot be simply approximated by the
expectation value of the Hamiltonian with respect to each state.
The mass and the corresponding eigenstate are obtained by diagonalizing
the energy matrix $\E$ defined as
\beq
\E_{ab} = (M_{sol} + \ve_n) \delta_{ab} + \frac{1}{2\I}
{^{}_a\!\ll i,j_\ell^\pi |(\vR - \vT(\infty))^2 | i,j^\pi_\ell \rr_b^{}}.
\hskip5mm(a,b=1,2,\cdots)
\eeq
We proceed with the $i$=1, $j_\ell^\pi$=$1^+$ state as an illustration.
The Wigner-Eckart theorem enables us to write down the expectation value
of $\vec\Theta$ with respect to the Fock state as
\beq
^{}_a\!\langle k'_\ell,m' | \Theta^q(\infty) | k_\ell,m\rangle^{}_b
=
\frac{\ts(\kl k_3 1 q|k'_\ell k'_3)}{\ts\sqrt{2k'_\ell+1}}
{^{}_a(k'_\ell\|\vec\Theta\|\kl)_b^{}}
\eeqlb{WigEck}
with the ``reduced matrix element" $(k'_\ell\|\vec\Theta\|\kl)$
and $\Theta^{\pm1}\equiv \mp\frac{1}{\sqrt2}(\Theta_x \pm i\Theta_y)$,
$\Theta^0 \equiv \Theta_z$.
With the help of eqs. (\ref{E1}b) and (\ref{WigEck}),
we obtain the energy matrix with
respect to the three states
$|1,1^+\rr_a(a=1,2,3)$\footnote{Actually, %----------- Footnote --------
we have two more states with $i=1$, $j_\ell^\pi=1^+$ made of {\em unbound}
heavy meson states $|0\rangle_3$ and $|2\rangle_-$ combined with
$i=1$ rotor spin state. We do not include these states into the
procedure, since they have large energy discrepancy with the other
three and their inclusion affects the eigenenergy only in the next
order in $1/N_c$.}%----------------------------------------------------
\beq
\E_{(1,1^+)} = M_{sol} - \half\gF + \dps\frac{11}{8\I} +
\frac{1}{4\I} \left(\!\!\bea{1}{ccc}
  -1   \!&\! 0   \!&\! \sqrt3 \\
   0   \!&\! 2   \!&\! 0  \\
\sqrt3 \!&\! 0   \!&\! 1
\eea\!\!\right).
\eeq
It leads us to three mass eigenvalues as
\beqa{1}{l}
M^-_{(1,1^+)} = M_{sol} - \frac12\gF + {7}/{8\I}, \\
M^+_{(1,1^+)} = M_{sol} - \frac12\gF + {15}/{8\I}, \mbox{ (doubly
degenerate)}
\eeqa
only the first of which is below the nucleon-heavy meson threshold.
The rotational energies are sufficiently large to make the other
states unbound.
In Table~3, listed are mass formulas for the positive and negative
parity $P$-baryon states obtained in a similar way. We present only
the states expected to be below the threshold. The degenerate mass of
the $P$-baryons with the same isospin and spin but different parity is
very interesting. However, we are not in the position to conclude
whether such a parity doubling has any physical importance or just an
artifact from our approximation on the radial functions.
\begin{table} %--------------------- Table 3 ----------------------
\begin{center}\as{1.5}
{\bf Table 3 } : Positive and Negative Parity $P$-baryon masses (in MeV).
\vskip 3mm
\begin{tabular}{ccccccr}
\hline
 $i$ & $j_\ell^\pi$ & $j^\pi$ & Mass Formula &
 $m^{}_{\Pc}$ &
 $m^{}_{\Pb}$
& \mc{1}{c}{b.e.$^*$} \\
\hline
 0 & $0^-$ & ${\frac12}^-$ &
 $M_{sol} + \m_{\Phi}^{} - \frac12\gF + 3/8\I$ & 2704 & 6042 & 210 \\
 0 & $1^-$ & ${\frac12}^-,{\frac32}^-$ &
 $M_{sol} + \m_{\Phi}^{} - \frac12\gF + 3/8\I$ & 2704 & 6042 & 210 \\
 1 & $1^-$ & ${\frac12}^-,{\frac32}^-$ &
 $M_{sol} + \m_{\Phi}^{} - \frac12\gF + 7/8\I$ & 2802 & 6140 & 112 \\
\hline
 0 & $1^+$ & ${\frac12}^+,{\frac32}^+$ &
 $M_{sol} + \m_{\Phi}^{} - \frac12\gF + 3/8\I$ &  2704 & 6042 & 210 \\
 1 & $0^+$ & ${\frac12}^+$ &
 $M_{sol} + \m_{\Phi}^{} - \frac12\gF + 7/8\I$ &  2802 & 6140 & 112 \\
 1 & $1^+$ & ${\frac12}^+,{\frac32}^+$ &
 $M_{sol} + \m_{\Phi}^{} - \frac12\gF + 7/8\I$ &  2802 & 6140 & 112 \\
\hline
&&\mc{4}{l}{$^*$ binding energy below the nucleon-heavy meson threshold.} \\
\end{tabular} \end{center}
\end{table} %--------------------- Table 3 end ----------------------

Though very rough, at this point, we may give a prediction on the
$P$-baryon masses. To this purpose, we add the heavy meson masses
$\m_{\Phi}(\equiv\frac14(3 m_{\Phi^*}^{} + m_{\Phi}^{})$,
the weight average of the heavy meson masses;
$\m_{D}$=1975 MeV and $\m_{B}$=5313 MeV)  to the mass formula.
Next, we fit the parameters $f_\pi$, $e$ and $g$ (equivalently,
$M_{sol}$, $1/\I$ and $\gF$) so as to yield correct
masses of the nucleon, Delta \cite{ANW} and $\Lambda_c$ \cite{HB}:
\beqa{1.2}{l}
m_N^{} = M_{sol} + {3}/{8\I} = 939 \MeV, \\
m_\Delta^{} = M_{sol} + {15}/{8\I} = 1232 \MeV, \\
m_{\Lambda_c} = M_{sol} + \m_{D} - \frac32\gF + {3}/{8\I} = 2285 \MeV,
\eeqa
which lead to
\beq
M_{sol}=866 \MeV, \hskip 5mm 1/\I=195 \MeV \and
\gF = 419 \MeV.
\eeqlb{gF}
Combined with the slope of the profile function
$F^\prime(0)\sim -690$ MeV (in case of the Skyrme-term-stabilized
soliton solution), eq. (\ref{gF}) implies that $g \approx -0.61$ which
is not far from that of the non-relativistic quark model ($-0.75$) and
the experimental estimation ($|g|^2\slt 0.5$). This set of parameters
yields a prediction on the $\Lambda_b$ and on the average mass of the
$\Sigma_{c}^{}$-$\Sigma^*_{c}$ multiplets, $\m_{\Sigma_c^{}}
(\equiv\frac13(2 m_{\Sigma^*_c}+m_{\Sigma_c}^{}))$ as
\beqa{1,2}{l}
m_{\Lambda_b^{}} = M_{sol} + \m_{B} - \frac32\gF +
{3}/{8\I} = 5623 \MeV, \\
\m_{\Sigma_c^{}} = M_{sol} + \m_{D} - \frac32\gF +
{11}/{8\I} = 2483 \MeV,
\eeqa
which are comparable with the experimental value of the $\Lambda_b$ mass
5641 MeV and $\Sigma_c$ mass 2453 MeV \cite{PDG}. The $P$-baryon masses
given in Table~3 could be accepted within the same error range.

However, all states listed in Table~3 do not seem to survive under the
finite heavy meson mass corrections. Recently, we have reported that such
finite mass corrections reduce the binding energy by an amount from
25\% (in case of bottomed baryons) to 35\% (in case of charmed baryons)
of their infinite mass limit, $\frac32\gF$ \cite{OPM}. Note that 35\%
of $\frac32\gF$ exceeds the binding energy $\frac12\gF$ for the
soliton-antiflavored heavy mesons. Such an ambiguity prohibits us from
any decisive conclusion before the finite mass corrections are
incorporated \cite{OPM2}. Nonetheless, the states with $i=0$,
$j_\ell^\pi=0^-$ and $i=0$, $j_\ell^\pi=1^\pm$ reveal a strong
possibility for the non-strange $P$-baryon(s) different from the quark
models. It supports the work of Riska and Scoccola \cite{RS93}, while the
binding energy and the mass formula are quite different from theirs.

\section*{Acknowledgement }
The work of YO was supported by the National Science Council of
ROC under grant \#NSC83-0208-M002-017. BYP and DPM were supported
in part by the Korea Science and Engineering Foundation through the
Center for Theoretical Physics, SNU.

%--------------------- Bibliography -------------------

\end{document}